# Magma oceans as a critical stage in the tectonic development of rocky planets

**Laura Schaefer[1,2,\*] and Linda T. Elkins-Tanton[1]**

[1]*School of Earth and Space Exploration, Arizona State University, Tempe, AZ 85287, USA*
[2]*ORCID: 0000-0003-2915-5025*

## Summary

Magma oceans are a common result of the high degree of heating that occurs during planet formation. It is thought that almost all of the large rocky bodies in the Solar System went through at least one magma ocean phase. In this paper, we review some of the ways in which magma ocean models for the Earth, Moon, and Mars match present day observations of mantle reservoirs, internal structure, and primordial crusts, and then we present new calculations for the oxidation state of the mantle produced during the magma ocean phase. The crystallization of magma oceans likely leads to a massive mantle overturn that may set up a stably stratified mantle. This may lead to significant delays or total prevention of plate tectonics on some planets. We review recent models that may help partly alleviate the mantle stability issue and lead to earlier onset of plate tectonics.

## Magma oceans are common

Magma oceans appear to be a common outcome of formation processes for large rocky bodies [1,2]. Small planetesimals and embryos can form on very short timescales in the protoplanetary disk, so that they incorporate significant amounts of short-lived radionuclides such as $^{26}$Al and $^{60}$Fe that can produce enough heat to at least partially melt many of these objects [3,4]. Models suggest that the asteroid 4 Vesta, which is the source of the howardite-eucrite-diogenite (HED) meteorites, went through a magma ocean stage due to such short-lived heating [5-7]. Many iron meteorites, which are likely the remnants of differentiated planetesimals, have ages of only ~1 Myr after Solar System formation, indicating very early, short-lived magma oceans [8,9].

*Author for correspondence (lschaefer@asu.edu).

Larger objects like the Earth take long enough to assemble that short-lived radionuclides cannot provide sufficient heating to melt them. For these larger, later-assembling bodies, substantial heating can come from accretionary impacts [e.g. 10,11] and gravitational segregation of metallic iron into the centre of the planet. Giant impacts also appear to be common in N-body simulations of rocky planet formation [12], and are likely to cause wide-spread melting that can lead to differentiation in both silicates and metals [13].

In this paper we will discuss several ways in which forward magma ocean models do a good job of matching real-world observations of the Earth, Moon, and Mars. Magma ocean models can produce large, low-shear velocity provinces (LLSVPs) at the base of Earth's mantle through cumulate overturn and predict that these structures are stable over billions of years. Lunar magma ocean models can produce the internal differentiation of the Moon, including the anorthositic crust, the source regions for the picritic glasses and mare basalts, and the KREEP component, rich in incompatible elements. Models of the magma ocean on Mars can produce early crust, some of which may be preserved to the present day, and which could be hydrated by the earliest atmosphere to produce primordial clays. We also introduce new calculations that show that magma ocean models can produce the present-day mantle oxidation state of the Earth through metal-silicate equilibration within the magma ocean, with no additional mechanisms required. We will then discuss the mantle state at the end of the magma ocean period and discuss whether it can hinder or allow early onset of plate tectonics.

# Magma ocean models match observations

Magma ocean models have been applied to all of the terrestrial planets, as well as the Earth's Moon. These models match observable properties of the present-day planets. The case for magma oceans on the Moon and Mars are easier to make because of the lack of active tectonic processing over most of the Solar System history. For the Earth, models can match deep mantle structures and the chemical properties of the mantle, including the abundances of trace elements and the present-day oxidation state. We review these model predictions below and present new calculations of the oxidation state of the magma ocean during core formation on the Earth.

## Differentiation of the Lunar Interior

Magma ocean models were first fully developed to explain observations and measurements of Apollo returned samples from the Moon indicating that there was abundant plagioclase feldspar in the lunar crust [14-16]. The feldspar found in the lunar crust is nearly pure (> 98%) anorthosite, containing trace amounts of mafic phases [15,17,18]. Lunar meteorites, remote spectroscopy and gravity field studies all support the wide-spread distribution of anorthosite in the lunar crust [19-21]. This widespread occurrence has led to the hypothesis that the Moon was at least partially molten and that anorthite plagioclase buoyantly segregated from the melt to form the crust [14-16]. This early silicate differentiation event is widely believed to have been a lunar magma ocean, which would be a natural consequence of the Moon-forming impact. In addition to the anorthositic crust, the existence of the highly incompatible element enriched KREEP (K, Rare Earth Element and P -rich) component, as well as the positive Eu anomalies in the highlands anorthosites and complementary negative Eu anomalies in mare basalts [e.g., 22-24] provide additional evidence in favour of a global lunar magma ocean. Models of metal-silicate partitioning during lunar core formation are also consistent with equilibration occurring at the lunar CMB, indicating a fully molten Moon [25].

Numerous authors have modelled the crystallization of the lunar magma ocean [e.g., 13, 26-33]. The lunar magma ocean would initially be fully molten at the surface and could rapidly cool and solidify by ~80% by volume in the first 1 ky [33]. The first solids would include Mg-rich olivine followed by orthopyroxene in the lower lunar mantle. Above 70-80% solidification, plagioclase begins to crystallize, and since it is lower in density than the magma ocean liquid, it should have floated to the surface to form a thermally conductive lid. Expulsion of trapped liquid by compaction and other processes would limit the amount of mafic minerals that could have formed in the crust, in agreement with observations [e.g., 34,35]. The magma ocean melt becomes more Fe-rich as crystallization proceeds, with a very dense ilmenite-bearing layer forming after about 95% crystallization. The residual liquids are very enriched in incompatible elements and have compositional characteristics similar to those found in the KREEP component [30]. Thermal models show that the conductive anorthositic crust (~40-50 km) would have delayed the solidification of the remaining magma ocean by ~10 My [33]. However, ages for lunar crustal materials, discussed below, suggest that magma ocean crystallization was not complete until ~150 – 200 Myrs. Recent models have shown, however, that tidal heating may extend the crystallization of this final magma ocean liquid out to 200 Myr [36,37]. Gravitationally driven cumulate overturn could produce later episodes of volcanism that would explain the ages and compositions of the Mg-suite of lunar crustal rocks [33].

Magma ocean crystallization experiments [32, 38, 39] reproduce many aspects of the numerical magma ocean models, including increasing density of the cumulates with crystallization, but some discrepancies remain. In particular, although liquid evolution trends are broadly similar between experiments and models, the late stage liquid composition can differ significantly [Gaffney et al. in prep]. Recent experiments suggest that a dry lunar magma ocean would produce much thicker anorthositic crust than observed by the GRAIL mission, but a wet lunar magma ocean containing 270-1650 ppm of water would produce a thinner crust [38,39]. This water abundance is consistent with recent measurements of water abundances in lunar anorthosites [40]. Additional models including the effect of water on the crystallization time will be needed to compare with lunar geochronology measurements discussed below in order to confirm a wet lunar magma ocean.

The age of lunar crustal anorthosites provides a constraint on the crystallization age of the lunar magma ocean, but measured ages cover a large span, from 4.47 to 4.29 Gyr [e.g., 41-46]. Kleine et al. [47] reviewed the isotopic ages for lunar samples available at the time and concluded that the combined constraints from the Hf-W and $^{147}Sm$-$^{143}Nd$ systems indicated that the lunar magma ocean reached ~70% crystallization, at which point anorthosite flotation should begin, no earlier than 60 Myrs and no later than 150 Myrs. Borg et al. [48] reported a very young age of crystallization of a ferroan anorthosite of 4360 +/- 3 million years from combined analysis of the $^{207}Pb$-$^{206}Pb$, $^{147}Sm$-$^{143}Nd$ and $^{146}Sm$-$^{142}Nd$ isotopic systems, making this the most reliable FAN age, and suggested that the lunar crust was produced by non-magma ocean processes such as serial magmatism. Boyet et al. [49] measured Sm-Nd systematics in several lunar ferroan anorthosite samples and found that they did not form a single isochron, which the authors suggest indicates that all but one of these samples formed through later magmatism or processing after the lunar magma ocean crystallization. The one sample with the largest deficit in $^{142}Nd$ gave a crystallization age of 60-125 Myr after Solar System formation for lunar crust formation. Recent results from Pb-isotopic evolution in lunar basalts has suggested a large differentiation event at 4376 ± 18 Myrs, which could coincide with the last stages of lunar magma ocean crystallization [50], consistent with the 142[Nd] ages for FAN 60025 [48]. Ages determined for lunar zircons indicate a younger age of 4.51 Gyrs for lunar crustal formation [51], which is significantly older than ages from ferroan anorthosites. This age, which is closer to Moon-formation ages of ~ 60 Myr from Hf-W [47], may instead simply represent the time of zircon formation in the initially rapidly cooling lunar magma ocean [52]. Modelling by Meyer et al. [36] suggests that prolonged cooling of the crust due to tidal heating could explain the late ages determined for the lunar anorthosites relative to the much shorter expected cooling times [e.g. 33].

One significant observation that cannot readily be explained by lunar serial magmatism, and therefore requires a deep magma ocean, is the apparent ubiquity of an orthopyroxene+olivine melting source region for the picritic glasses between ~250 and ~500 km depth on the lunar nearside (data summary in [33], Figure 6). This thick deep layer of uniform mineralogy has to have been created by a deep, fractionating magma ocean: Serial diapirs will not leave behind a uniform cumulate pile, and the heterogeneities from between the putative melting regions would be enriched and therefore melt at lower temperatures later, producing melt products different from the picritic glasses, and not sampled on the Moon.

Further, the lunar crust's thickness is smoothly varying from the nearside to the far side, excepting the effects of large impacts. To produce a smoothly varying crustal thickness via serial magmatism, the wavelength of the diapirs must not be more than a few multiples of the crustal thickness, or relaxation of the rapidly-cooling and high-viscosity anorthosites would not produce a smoothly varying crustal thickness. Such ubiquity of melting around the Moon must be virtually indistinguishable from a global magma ocean.

Many aspects of the lunar magma ocean remain to be investigated, including the depth of the magma ocean, partial equilibrium vs fully fractional crystallization [30,33], the cooling times of the crust and the possible formation of a crustal dichotomy between near and far-side of the Moon [53, 54]. While many details of the chronology of lunar geology remain to be explained, the lunar magma ocean model is very successful in producing many of the observed properties of the Moon's internal structure and surface composition.

## Martian crust and clay formation

An early Martian magma ocean is supported by measurements of short-lived radioactive isotopes that indicate early differentiation of Mars and generation of enriched and depleted mantle reservoirs [9, 55-57]. Young core formation ages for Mars (>~4-10 Myrs) [58] are further argument for a magma ocean because short-lived radioactive isotopes such as $^{26}$Al would still have been present to provide additional heat and core segregation requires at least partial melting of the silicate mantle. However, it remains uncertain how deep the magma ocean on Mars would have been. Physical models of the accretion and differentiation of Mars have attempted to determine the degree of melting with variable results. Models using impact heating and cooling but no other heat source find only partial melting of the near surface and a cold interior [59]. A multiphase model including radioactive and gravitational heating found that core formation may be a catastrophic, early event that can lead to additional heating of Mars-sized objects, although full silicate melting is not required [60]. The blanketing effect of a hybrid proto-atmosphere (both outgassed and nebular components) could also help to sustain a global magma ocean throughout accretion, depending on the accretion rate of proto-Mars [61]. No model including all of the physical processes controlling the heating and cooling of proto-Mars (impact heating & cooling, radioactive decay, atmospheric blanketing, gravitational segregation, thermal convection, etc.) exist, and the physical partitioning of energy of those processes remains largely unknown, so the depth of the martian magma ocean remains uncertain.

The magma ocean stage may have produced silicate reservoirs enriched and depleted in incompatible elements during fractional crystallization, which remain today due to incomplete mantle mixing. The ages of these reservoirs as measured with short-lived radioactive isotope systems have been used to infer the duration of the magma ocean stage and timing of crust formation, although interpretations within the community can vary. Measurements of Sm-Nd isotope systematics in shergottites by [62] suggest that the magma ocean stage on Mars lasted for ~100 Myrs, with a depleted reservoir forming by 4.535 Gyr and an enriched reservoir, plausibly either the crust of late stage magma ocean liquids, forming by 4.457 Myrs. Lead-lead ages of zircons in the polymict breccia NWA 7034/7355, which formed through melting of primary crust, give maximum crustal formation times of 4.428 Gyr [63]. Whole rock Sm-Nd ages of the same meteorite give similar ages of 4.42 - 4.46 Gyr [64]. Co-existing variations of $\varepsilon^{182}$W and $\varepsilon^{142}$Nd have been attributed to multiple episodes of silicate differentiation [9]. These authors find formation times of the depleted reservoir of ~25 Myrs, with ongoing formation of the enriched reservoir (thought to be the crust), extending to at least 40 Myrs after Solar System formation. This is ~60 Myrs younger than [62] and ~100 Myrs younger than [63, 64] but is more consistent with results from martian crust formation models [65]. In comparison, magma ocean thermal models predict crystallization timescales < 5 Myr for Mars [66, 67]. Additional analyses of Sm-Nd isotope systematics in shergottites and nahklites found a silicate differentiation time of 4504 Myr (~60 Myrs after Solar System formation) [68]. Given the discrepancy with magma ocean thermal models [66, 67], these authors suggest that this age could be explained by a giant impact and heterogeneous mixing of the martian mantle, rather than a global silicate differentiation event. However, recent high-precision isotopic work on zircons by [69] shows that the Martian magma ocean was solidified in less than 10 Myr after planetary formation, a result that indicates that a solid Mars existed within 20 Myr of CAIs. This result supports the rapid solidification of the Martian magma ocean and indicates that later reservoir disruptions were likely due to giant impacts and remixing.

Elkins-Tanton et al. [65] modelled early crust formation on Mars at 30-50 Myrs after accretion arising during cumulate overturn by melting of the hot cumulates rising into the upper mantle. Measurements of potentially ancient crust in crater central peaks find primarily dunite and pyroxenite-dominated lithologies.

Compositional variations are suggestive of high degrees of fractional crystallization in a slowly cooling magma body, potentially generated by the overturn of the mantle following magma ocean solidification [70]. However, [71] find that stagnant lid convection models beginning from a thermal and density state derived from fractional crystallization cannot reproduce the later long-term thermal evolution of the planet (e.g. 1-2 long-lived mantle plumes). Therefore, more work is still needed in understanding the depth and longevity of the martian magma ocean, and the post-overturn thermal state.

Clays are relatively common on the surface of Mars [72], suggesting the widespread presence of water on the surface (or near-surface) throughout the Noachian. Their presence has been used to try to constrain the climate history of early Mars, with models ranging from 'warm and wet' to 'cold and transiently wet' [73,74]. However, magma ocean models provide another possible source for clay minerals. Magma ocean outgassing would result in a thick atmosphere, of which a large or primary component, is water vapor [e.g. 66]. Primitive clays found on the surface of Mars could have formed through rapid reaction of a supercritical steam atmosphere or hot early ocean during the late magma ocean stage with the earliest magma ocean crust [75]. Although this clay layer would later be buried by new crust [66], this model may account for the presence of a thick clay layer at large depths within the Martian crust observed in crater central peaks [76] and the abundance of clay minerals within the regolith. The clay formation model assumes the long-term stability of the crust against the overturn, which should be further investigated. Early clays would likely have lower D/H ratios, since they would have formed before significant water escape from the planet, which provides a testable prediction of the primordial clay model.

## Chemical Heterogeneities in the Earth's Lower Mantle

Seismic studies have long noted large scale heterogeneities in the lower mantle from variations in seismic velocities [77,78]. Both seismically fast and slow regions have been identified. While the seismically fast regions seem to be associated with regions of subduction, the seismically slow regions --called large low shear velocity provinces, or LLSVPs -- are found in regions associated with hot spot volcanism [79,80]. Seismic tomography suggests that the LLSVPs are associated with chemical heterogeneity in the lower mantle, and that these regions have higher densities than the surrounding lower mantle [81,82]. Within the LLSVPs are small (5-40 km thick) regions of even lower seismic velocity (ultralow-velocity zones, ULVZs), that have been associated with partial melting of silicates directly above the core or with significant compositional differences [83 -86]. Geodynamics simulations suggest that the LLSVPs could have been stable throughout Earth's history if they are higher in density and/or viscosity than the surrounding lower mantle [87-89], suggesting a potential primordial origin.

Chemical heterogeneities within the mantle suggest the existence of a small enriched mantle reservoir in the deep mantle [43, 90]. This suggestion is based on the Earth's $^{142}Nd/^{144}Nd$ ratio, which is higher than measured in chondrites and suggests that an early silicate differentiation event occurred in the first 160 Myrs [91]. Other recent measurements [92] suggest that the Earth has the same $^{142}Nd/^{144}Nd$ ratio as enstatite chondrites, obviating the need for an enriched reservoir. Other arguments in favour of a primordial chemical reservoir in the mantle come from noble gases. Mid-ocean ridge basalts (MORBs) and ocean island basalts (OIBs) appear to tap different chemical reservoirs, with the MORB source being depleted in primordial noble gas components and the OIB source being enriched and potentially undegassed [93-97].

Fractional crystallization of a magma ocean produces an unstable density stratification due to the enrichment of Fe in late stage melts [1, 66, 98-100]. These models predict mantle overturn following the crystallization of the magma ocean, with the final iron-rich residues sinking to the core mantle boundary of the planet. Since they are denser than the overlying mantle, these overturned cumulates might be stable at the CMB on long time scales [e.g. 101], so models suggest that these could be responsible for the modern day LLSVPs [102].

Alternative primordial origins for the LLSVPs have been suggested, including a primary basal magma ocean [103], partial melting of the Archaean mantle and sinking of dense, Fe-rich melts [104], or subduction of early crust enriched with late delivered chondritic material [94105 A primary basal magma ocean is produced if the liquidus and the magma ocean adiabat first intersect in the middle of the mantle, with crystallization proceeding towards the surface rapidly, and more slowly towards the core. The basalt melt would represent an un-depleted primordial chemical reservoir, which could generate denser crystallization products as Fe becomes enriched in the melt [103]. Zhang et al. [106] alter the basal magma ocean model to suggest that the higher density of the LLSVPs is due to the trapping of Fe-Ni-S metallic liquids within the basal magma ocean cumulates. However, recent analysis of mantle melting curve [107 ] suggests the mantle liquidus should increase smoothly from surface to core-mantle boundary (CMB). If this is verified by further experiments, magma ocean crystallization would initiate at the CMB, and a primary basal magma ocean could likely be ruled out.

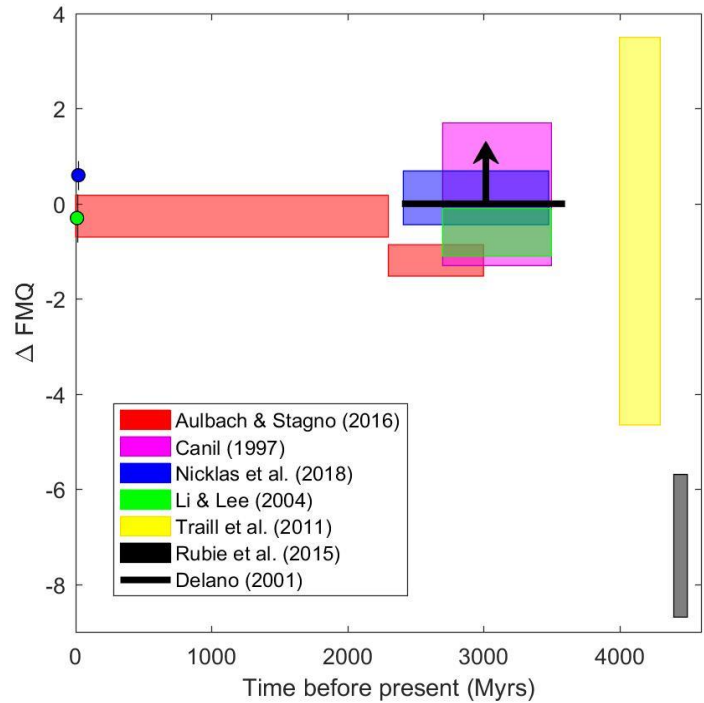

## Figure 1

Oxidation state of Earth's mantle through geologic time from redox proxy measurements. See text for source details. The two data points at ~ 0 Myrs are values for present day MORBs. We plot historical oxygen fugacity relative to the fayalite-magnetite-quartz buffer (FMQ). At 1200C and 1 GPa, NNO = FMQ + 0.7, IW = FMQ − 3.69.

## Mantle oxidation state

The oxidation state of the upper mantle of the present-day Earth is near the FMQ buffer [108]. Various estimates are of the mantle redox state through time are shown in Figure 1. Measurements of ancient redox proxies, Cr and V whole-rock abundances in ancient volcanics and the composition of Cr-rich spinels in ancient volcanics indicate that the oxygen fugacity of the mantle has been at or above the present-day value ($\pm 0.5$ log $fO_2$) since ~3900 Ma [109]. Canil [110] used partitioning of V between komatiitic liquids and olivine in 6 well-characterized komatiite flows to estimate Archaean $fO_2$ and found values from $\Delta$NNO -2 to +1. Li & Lee [111] looked at V/Sc ratios in basalts and find that Archaean basalts have V/Sc ratios only slightly smaller than that of modern MORBS, indicating only a minor decrease of ~0.3 log $fO_2$ units relative to FMQ. Recent measurements of 3.5 – 2.4 Ga komatiites used V partitioning behaviour to constrain the $fO_2$ of the mantle, and found a range of -0.11 to +0.43 FMQ, consistent with a relatively constant mantle $fO_2$ through time [112]. Aulbach & Stagno [113] used V/Sc ratios in metamorphosed mid-ocean ridge basalts (MORBs) and picrites from 3.0 – 0.55 Ga as redox proxies and found a slightly larger decrease in mantle $fO_2$, down to -1.5 FMQ, during the Archean. They note that the increase of the mantle $fO_2$ to the present-day value appears coincident with the atmospheric Great Oxidation Event (GOE). Analysis of Ce abundances in detrital Hadean-aged Jack Hills zircons suggest oxygen fugacities consistent with the present day back to 4.3 Ga [114]. Taken together, the present measurements are consistent with a constant oxygen fugacity within the mantle over the entire history of the Earth.

However, models of core formation, which occurs within the magma ocean of the forming Earth, require much more reducing conditions in order to match trace element partitioning behaviour between the mantle and the core [e.g. 115-117]. Therefore, oxidation of the mantle to the present-day state had to occur contemporaneously with core formation or shortly thereafter and operate on a relatively short timescale. Several models have been proposed to increase the oxidation state of the mantle through the post-production of $Fe^{3+}$ in the mantle. These include gradual H escape from accreted water [118], and the production of $Fe^{3+}$ through disproportionation of $Fe^{2+}$ in the lower mantle by crystallization of bridgmanite [119,120]. Trace element partitioning models [e.g. 117, 121] match the present-day mantle FeO abundance by allowing the composition of accreting material to become more oxidized through time (IW-5 to IW-2), but these models fail to predict $Fe^{3+}/\Sigma Fe$, which governs present-day oxidation state. These models typically end with a mantle at IW-2, about 5 log $fO_2$ units below the present day.

We propose an alternative mechanism: that disproportionation of $Fe^{2+}$ occurs in the silicate melt phase in the magma ocean during equilibration with the core forming metal delivered by accreting impactors, rather than during crystallization. This allows immediate separation of the metallic liquid, avoiding the remixing problem of [119,120]. The equilibrium amongst the different Fe valence states will be given by the reaction $3FeO(melt) \rightarrow 2FeO_{1.5}(melt) + Fe$ (liquid metal). The metal will sink through the magma ocean to join the core, leaving the oxidized mantle material behind, without requiring complicated scenarios of melting and recrystallizing of bridgmanite that would be difficult to quantify. To reproduce the present day upper mantle, metal-silicate equilibration in the melt at high pressures must produce $Fe^{3+}/\Sigma Fe \sim 0.01 – 0.05$ [108]. As in [122], we assume that equilibrium will be controlled at the level of metal-silicate equilibration and the magma ocean

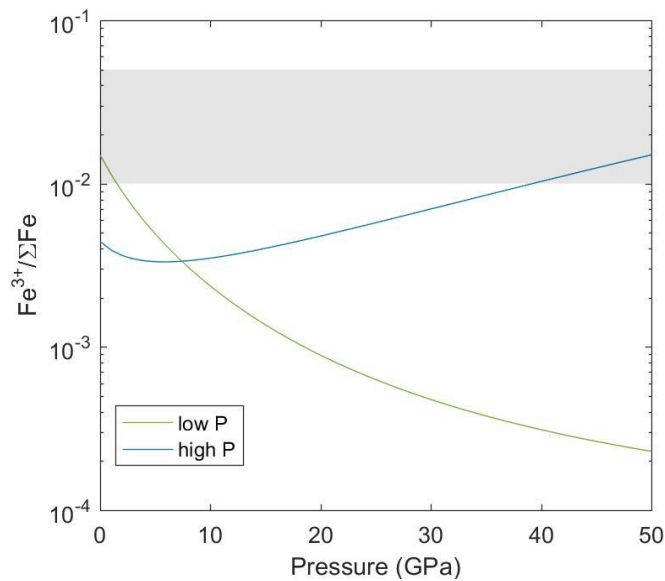

## Figure 2

Fe³⁺/ΣFe in the magma ocean produced by equilibration of liquid silicate and core-forming metal as a function of pressure along the mantle liquidus of [123]. See Table 1 for thermodynamic data sources. The grey shaded region highlights the Fe³⁺/ΣFe values in the Earth's present day upper mantle [108].

composition would be rapidly homogenized by convection. In Figure 2, we show the amount of $Fe^{3+}$ produced through this reaction as a function of temperature along the mantle liquidus [123], using different data sets for partial molar volumes of the molten iron oxides [124, 125]. The low-pressure model data [124, 126] was measured at 1 bar, whereas for the higher-pressure model, we use data from new measurements of the Fe³⁺/ΣFe ratio in high pressure melts [125] of up to 7 GPa to refit the partial molar volume of the $Fe^{3+}$ oxide. We use the equation of state of liquid Fe metal from [127] for all calculations and include partitioning of oxygen and silicon into the metal phase using the partitioning model of [128]. The thermodynamic data used for these calculations are given in Table 1. We calculate the Fe³⁺/ΣFe ratio as a function of pressure up to 50 GPa, which is close to average conditions of core formation determined for the Earth from trace element partitioning studies [117,121].

The updated equation of state for the liquid iron oxide from [125] predicts much larger $Fe^{3+}$ abundances are produced at high pressures than the 1 bar data of [124] do. This is consistent with new,

## Table 1.

Fit parameters used to calculate the $Fe^{3+}/\Sigma Fe$ ratio as a function of pressure (see Figure 2). We refit the data of [125] to a Murnaghan equation of state for $FeO_{1.5}$ (liq), while taking values from [124] and [126] for FeO(liq). For both models, we take the standard Gibbs energy of formation of the liquid oxides from [148] and calculate activity coefficients for the liquid oxides from [149]. For both models, we take a value of $\kappa' = 4$ for both oxides.

|  | Low P model | Reference | High P model | Reference |
|---|---|---|---|---|
| $\bar{V}_{FeO}$ (cm³/mole) | 13.65 | [124] | 13.65 | [124] |
| $\bar{V}_{FeO_{1.5}}$ (cm³/mole) | 21.065 | [124] | 18.9 | Fit to [125] |
| $\kappa_{FeO}$ (GPa) | 30.33 | [126] | 30.33 | [126] |
| $\kappa_{FeO_{1.5}}$ (GPa) | 16.6 | [126] | 10.8 | Fit to [125] |

unpublished measurements of $Fe^{3+}/\Sigma Fe$ in silicate melts from [129], which go up to pressures of ~25 GPa. The stabilization of $Fe^{3+}$ in the melt phase at high pressure was predicted by [122], partially on the basis of the stability of $Fe^{3+}$ in the crystalline phase at high pressures. As Fig. 2 shows, the equilibration of the different Fe valences at high pressure during the separation of core forming metal can easily produce the observed present day $Fe^{3+}$ abundances in the Earth's upper mantle, shown by the grey shaded region. Therefore, no additional oxidation mechanisms are required to produce the present-day oxidation state, which implies that there has been limited oxidation of the Earth's upper mantle over geologic time. The calculations with the new data also suggest that the mantles of larger planets should in general be more oxidized than smaller planets once the metallic phase has separated, given that more $Fe^{3+}$ is produced at higher pressures. This is corroborated by the low initial oxygen fugacity of the depleted martian mantle (~IW) [130], in comparison to the Earth's much higher oxygen fugacity. This has significant implications for the compositions of outgassed atmospheres we should expect on rocky exoplanets of different sizes.

# Magma oceans set the initial conditions for planetary tectonics

## Density stratification and magma ocean overturn

Models of fractional crystallization of magma oceans result in unstable density structures due to the enrichment of late-stage melts in iron. These crystallization models predict whole-mantle overturn at the end of magma ocean crystallization, which results in a stably stratified density structure in the earliest solid mantle. This stable stratification is predicted to inhibit the onset of whole mantle convection, which may result in an early stagnant lid tectonic regime on rocky planets [131, 132]. Recent models [133, 134] investigate the possibility of solid state convection beginning within the cumulate pile before the magma ocean has fully crystallized. Ballmer et al. [133] investigate incremental overturns within the solidifying cumulate pile using 2D numerical convection simulations and find that the majority of the mantle becomes well mixed. However, the final Fe-rich cumulate melt layer sinks to the CMB during the final overturn and remains stable as a highly dense layer for geologic time, which may be related to present day LLSVPs. Maurice et al. [134] similarly find that early onset of solid state convection while the magma ocean is still crystallizing will more efficiently homogenize the mantle than late stage overturn. However, their model fails to produce a LLSVP-like zone at the base of the mantle, and they instead prefer the basal magma ocean model of [103]. Tikoo & Elkins-Tanton [135] show that water within the late-stage melt may be expelled during mantle overturn, which would cause a large influx of water into the upper mantle. The large abundance of water in the early upper mantle would have allowed partial melting and lowered the viscosity, possibly allowing the early initiation of whole mantle convection. These models show that the mantle structure at the end of magma ocean phases remains uncertain.

## Tectonics after the magma ocean

Recent work in scaling theory, steady-state simulations, and numerical evolutionary models show that the initial conditions of a solid-state mantle convection simulation can strongly influence the tectonic style of the simulated planet [136-138]. For instance, [136] demonstrate that tectonic models show a history dependence, i.e. the tectonic mode of a system is determined by its specific geologic and climatic history. Therefore, the density and thermal structure at the end of the magma ocean stage and following cumulate overturn sets the stage for the future tectonic evolution of the planet. As discussed in the previous section, a number of parameters remain poorly constrained by the end stage of magma ocean models, which influences the later tectonic models. The uncertain parameters include the mantle potential temperature and heat flux, the degree of mantle hydration, the thickness of the lithosphere, and the planetary climate.

After discovering remnant magnetism on Mars, models predicting early plate tectonics were popular to explain rapid core cooling needed for magnetic field generation [139]. However, evidence of significant mantle heterogeneities on Mars suggest that significant convective mixing has not occurred. There is also little evidence of crustal recycling, indicating that Mars likely entered a stagnant or sluggish-lid phase following the magma ocean [71, 132], with heat fluxes out of the mantle dominated by a few long-lived mantle plumes.

*Phil. Trans. R. Soc. A.*

It remains uncertain if plate tectonics on the Earth can initiate immediately after the magma ocean stage, or if the magma ocean is more likely to be followed by a stagnant-lid mode [140] or alternative tectonic regime (e.g. squishy lid [141], heat pipe model [142], etc.). Foley et al. [143] investigate onset of subduction during the Hadean using a model including grain-damage and find that a period of stagnant lid convection follows the initial mantle overturn, and that sluggish subduction can initiate relatively rapidly, due to damage build-up in the cold, dense lid material. Recent models have highlighted the potential role of triggers for initiating subduction, including external impacts that punch through the lithosphere and mantle plumes [144, 145]. Numerical models of core-formation suggest that metal diapirs may set up conduits to permit early onset of mantle plumes, either purely thermal or thermochemically driven [146, 147], which is particularly effective if the magma ocean is restricted to the upper mantle. These early mantle plumes could then play a role in triggering onset of early subduction.

This discussion highlights the uncertainty in the tectonic state of planets following the magma ocean stage and the need for additional modelling. Magma ocean models must in future be more coupled with petrologic data from short-lived isotopic systems to help us better understand the early evolution of the terrestrial planets and the requirements for the onset of plate tectonics. Plate tectonics is often thought to be essential for the sustainability of life, and therefore such work has implications for the search for life on exoplanets as well as in our own Solar System.

# Additional Information

**Data Accessibility**
The dataset necessary to reproduce Figure 2 have been uploaded as part of the Supplementary Material. Sources of data shown in Figure 1 are given in the text.

**Authors' Contributions**
LS and LTET conceived of and designed the study, and LS drafted the manuscript. All authors read and approved the manuscript.

**Competing Interests**
The authors declare that they have no competing interests.

**Funding Statement**
The authors acknowledge funding from Arizona State University's Interplanetary Initiative, the NASA Psyche mission and the NSF CSEDI program.

*Phil. Trans. R. Soc. A.*